# Generating Forms via Informed Motion, a Flight Inspired Method Based on Wind and Topography Data

Demircan TAŞ
Osman SÜMER
...
(Istanbul Technical University, Institute of Science and Technology, Architectural Design Computing Graduate Program, demircan.tas@itu.edu.tr, sumer16@itu.edu.tr)

**Abstract.** Generative systems are becoming a crucial part of current design practice. There exist gaps however, between the digital processes, field data and designer's input. To solve this problem, multiple processes were developed in order to generate emergent and self-organizing design solutions that combine the designer's input with surface models acquired via photogrammetry and generative design tools. Different generative design methods were utilized for trials, including surface scattering based on UV coordinates, animation snapshots (similar to long exposure photography) and a particle swarm algorithm on arbitrary data, interpolated within GIS software. A large volume of adaptive forms that are complex, yet responsive to changes in parameters, user input, topography and/or various spatial data were acquired. Resulting outputs were rendered and projection mapped onto the original physical model and evaluated for further iterations.

## 1. INTRODUCTION

The use of computational techniques has shifted the paradigm of architectural design practice. Oxman [1] suggests that dynamic, responsive space and form, producing new classes of designs have been made possible by digital technologies. Design is no longer constrained to discrete shapes in the traditional sense. Design experimentation is not reliant on paper-based sketching, thanks to the exploitation of generative-based processes of transformation. Design context of the modernist approach may be iconic, stylistic or configurative while digital design context is a "per-formative shaping force acting upon shape, structure and material".

According to Carrara [2], architectural design processes can be defined in three steps. The definition of goals, generation of various solutions, and evaluation of the satisfaction of said goals provided by the suggested solution. Goals, generated through analytical and deductive means are transformed into structured sets of requirements. Created through intuitive and inductive processes, solutions are invented or adapted to contain performance characteristics that achieve the requirements and other desired attributes. While the solutions are evaluated, requirements are also modified in order to accommodate emerging opportunities as well as to overcome unconformable conflicts.

### 1.1. Problem Definition

Generative systems have been utilized for various phases of design, creating form via shape grammars, cellular automata, genetic algorithms, l-systems and swarm intelligence [3].

The subject of this paper is design work carried out in order to create a design algorithm based on generative systems, viable for the rapid generation of a large body of solutions to design requirements of complex nature. The intention in creating such algorithms was to create an explicit and responsive workflow that generates form, based on contextual information and fluent user input of both digital and analog nature.

Our work aims to combine digital algorithms with material design models and collected data. This paper aims to clarify our processes and trials, resulting in the creation of a swarm intelligence (SI) based generative design algorithm as part of a team effort within the Generative Systems in Architectural Design class within our master's studies while trying to make the "relentless boundary between the digital and the material" [4] more permeable.

## 1.2. Aims and Objectives

The main objective of our work is to reduce the creative gap between physical models and digital systems. Our aim is the creation of adaptive algorithms that generate initial design shapes based on contextual data and analog models. Arbitrary wind speed and direction data were utilized within our trials, yet data of different nature and origin can be fed into the system in order to create complex, contextual forms for differing environments and design goals. Another aim of this study is to create a coordinated [5] design algorithm where results are based purely on the input of design and field data. Ambiguity and experimentation are based on user interactions, not random variables.

Figure 1. Conversion of spreadsheet data to 2d spatial coordinates

| | A | B | C | D | E |
|---|---|---|---|---|---|
| 1 | Station | UTMX | UTMY | Direction | Speed |
| 2 | ST1 | 615380 | 9531496 | 22.5 | 16 |
| 3 | ST2 | 616004 | 9533157 | 45 | 14 |
| 4 | ST3 | 617630 | 9533808 | 90 | 12 |
| 5 | ST4 | 619191 | 9532326 | 135 | 12 |
| 6 | ST5 | 621295 | 9530663 | 112.5 | 7 |
| 7 | ST6 | 622303 | 9533906 | 180 | 6 |
| 8 | ST7 | 616992 | 9532169 | 10.5 | 18 |

Results of our trials will be compared with regards to the presence of emergence and self-organisation. De Wolf and Holvoet [6] define emergence as follows:

> "A system exhibits emergence when there are coherent emergents at the macro-level that dynamically arise from the interactions between the parts at the micro-level. Such emergents are novel w.r.t. the individual parts of the system."

Self-organisation is defined by the same authors as:

> "... a dynamical and adaptive process where systems acquire and maintain structure themselves, without external control."

Another aim of our work is to bridge the gap between the analogue material, created via human input -or taken from nature- and the digital, especially the generative tools provided via digital computation as a possible way to create and design with the digital, as opposed to the current paradigm of merely producing through the digital [7].

## 2. METHODS

Our workflow is based on the scanning of a physical model via photogrammetry tools and enriching the acquired 3d model with spatial data through raster maps. The acquired landscape model is combined with the raster data by the use of UV mapping techniques. Generative algorithms were tested on the informed digital model along the course of three trials, providing us with a volume of forms that can be utilized for design.

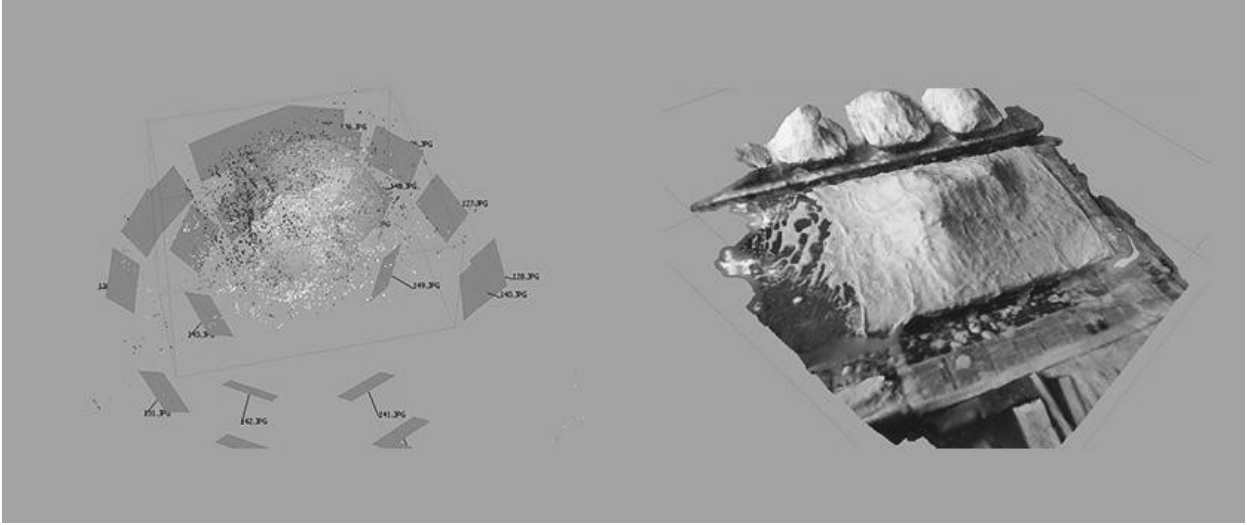

Figure 2. Digital landscape model acquired via photogrammetry

For the creation of the physical model, pieces of crumpled paper were used to generate underlying forms with the inherent characteristics of the material combination. Wet, plaster soaked bandages were placed on the paper base and left to dry. Multiple such topographies were constructed in order to test the adaptive nature of our system.

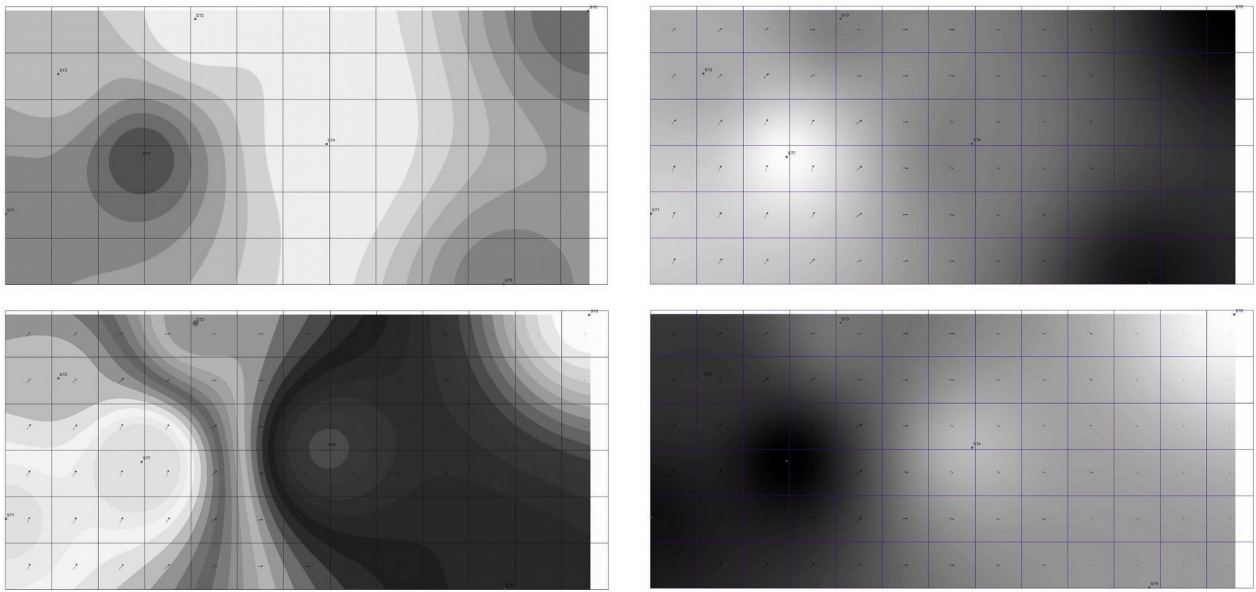

Figure 3. Raster maps acquired by interpolation

An off-the-shelf approach was adopted for most of the digital methods. Agisoft Photoscan software was used for obtaining digital models from the topographic models by digital photography. Animation software from Autodesk was utilized to retopologize the Triangulated Irregular Mesh (TIN) geometry to quad based representations in order to obtain higher quality surfaces for digital calculations.

The approach was tested on a fibre and moulding plaster composite, handmade landscape model which was digitized by means of photogrammetry. Arbitrary numbers were constructed as a simple database in spreadsheet software. Raster maps of continuous and differentiable nature were necessary for the application of the necessary tools on the digital surface. Data was converted into digital material of the necessary attributes by the use of inverse distance weighting algorithm within ArcGIS [8]. The interpolated raster data was mapped to the scanned surface via the use of UV [9] mapping techniques within Maya. Algorithms were iterated on the terrain geometry, connected to the spatial data within the animation software.

Responsive geometries were created with the use of digital animation techniques. These forms were instantiated on the landscape based on different methods with the aim of combining their adaptive nature with the underlying landscape.

Initially, multiple models of avian inspired forms were digitally modelled as polygonal forms. We have combined these models with a morphing tool, where one shape can be animated to transform into a different shape with similar polygonal topology.

Our second modeling approach is similar to long exposure photography [10] where the position of an object on different times is captured on a single image. Different simple geometries were animated based on wing motion of birds via an Inverse Kinematics (IK) algorithm [11]. Resulting positions within a certain timeline were combined in order to create static forms. Forms created through this method resemble the aesthetic characteristics of avian creatures.

Morphing and IK algorithms were utilized for different trials. Pieces were instantiated on a surface, driving the morph animation and orientation of each part via the field data. Motion of individual parts (via IK), resulting in the animation of the whole was instantialized for each frame, in order to acquire dynamic forms.

## 2.1. Trial I

Our first trial was based on morphing shapes. Multiple starting positions were created based on avian forms and a morph algorithm was utilized in order to interpolate any number of hybrid shapes.

The resulting shapes were placed on the terrain via a parametric scattering algorithm in a grid-like pattern. Affine transforms of each instance were exposed in order to match such attributes to data. Additionally the morph animation was limited within certain frames and frame number attribute was also exposed as an input channel.

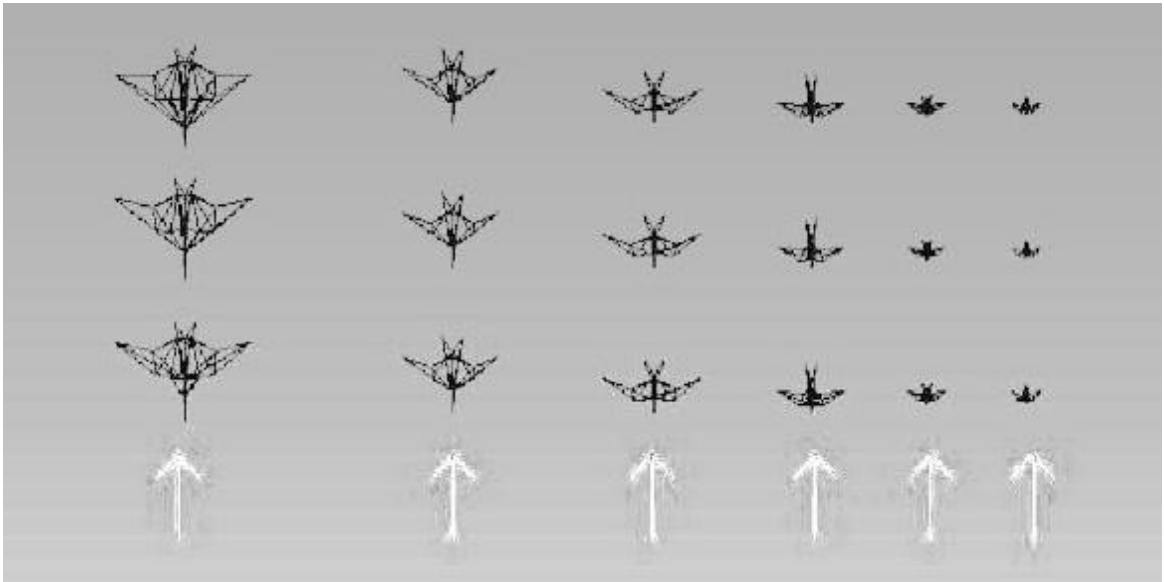

Figure 4. Multiple forms with similar topologies defined as morph targets

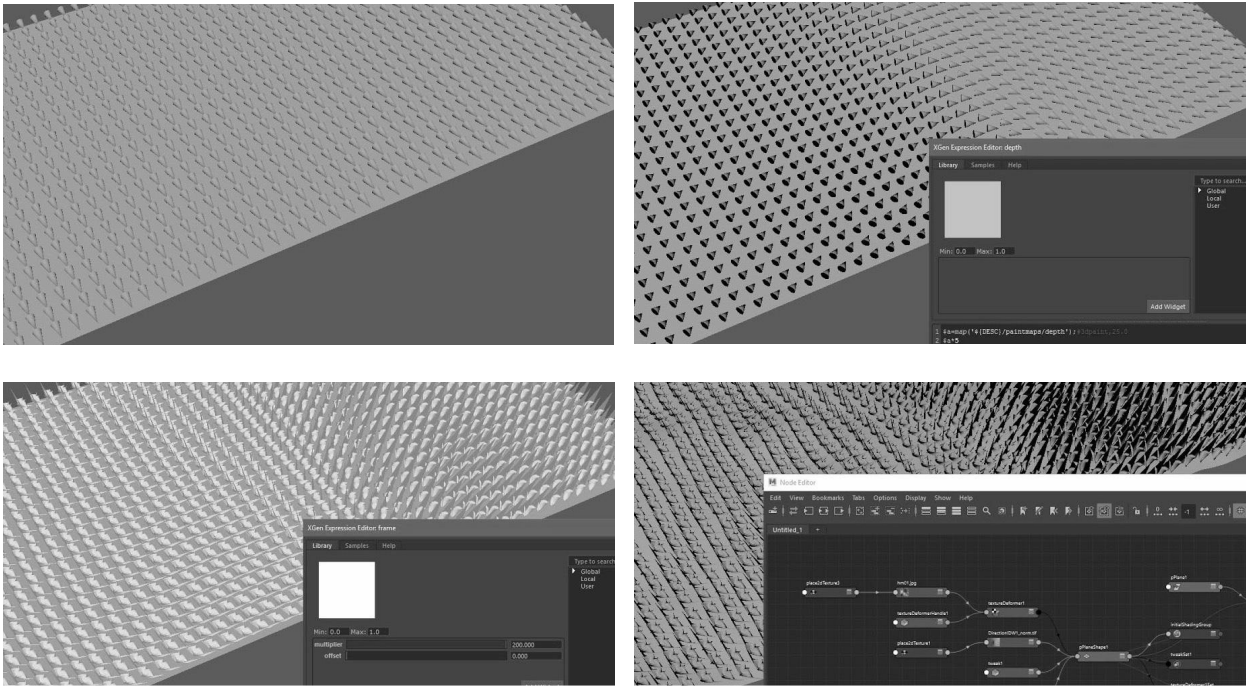

Figure 5. Progression of shapes, from regular arraying without data input through affine rotation, scale and frame connections

The linear grayscale raster images were used as input in Maya to provide local values to be input for each model instance. This resulted in each model acquiring its frame number from the wind speed, and orientation from the wind direction values. While initially regular and static, manipulation of local pivot positions caused dynamic patterns to emerge.

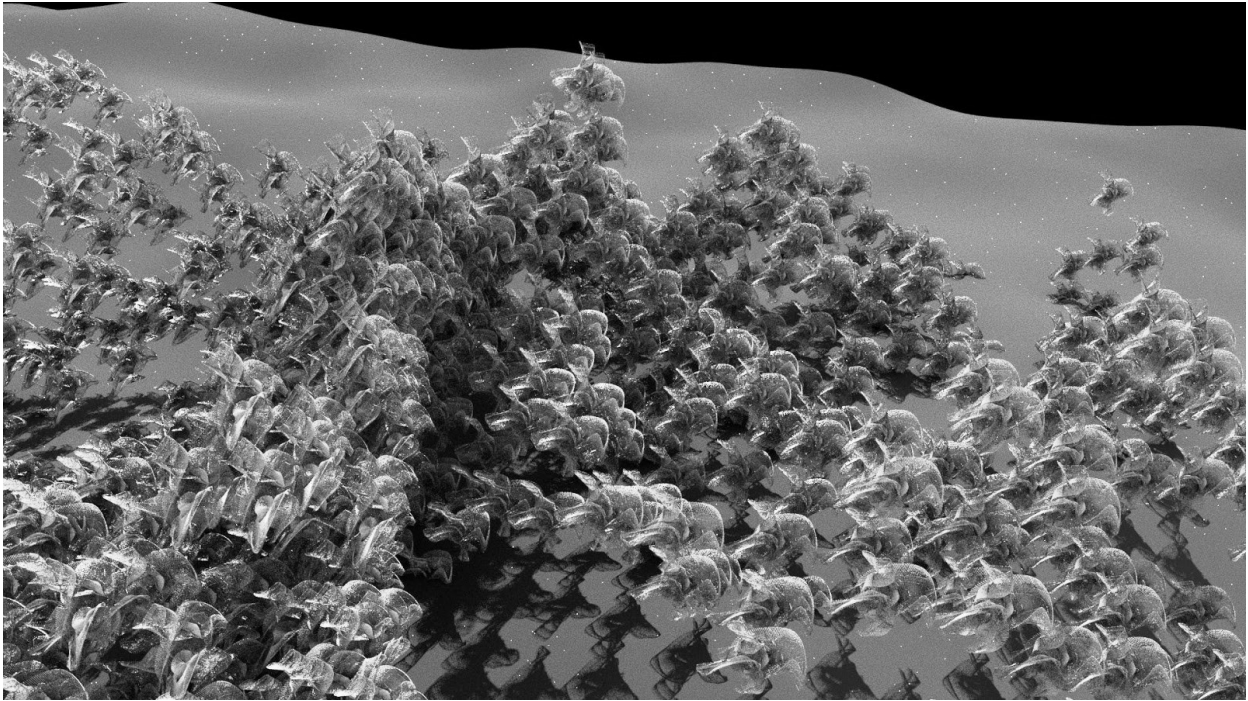

Figure 6. Emergence obtained via pivot point manipulation

## 2.2. Trial II

In addition to morph animation, we searched different results through an IK based method. A linear set of joints were created to be combined with an end locator (IK handle). By constraining the end locator to simple curve geometries, complex avian-inspired forms emerged from simple sweeping motions.

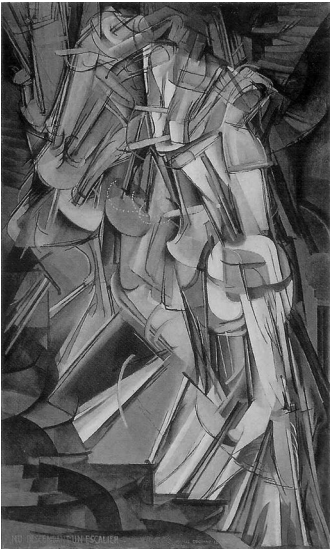
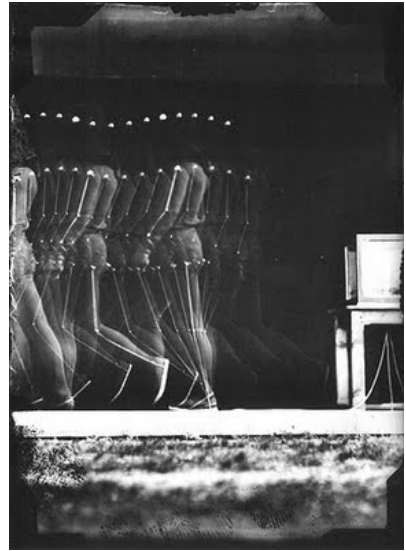

Figure 5. Marcel Duchamp, Nude Descending a Staircase, No. 2 (1912) Philadelphia Museum of Art, Louise and Walter Arensberg Collection (Left)

Etiene Jules Marey, Man Walking, 1890-1891 (Right)

Rather than recursively scattering the results as static models, motions were preserved and combined with an additional layer of animation. This resulted in more dynamic shapes that also involved characteristics of the given motion. Combination of different animations provides freedom when creating forms, even making it possible to draw with the top layer animation.

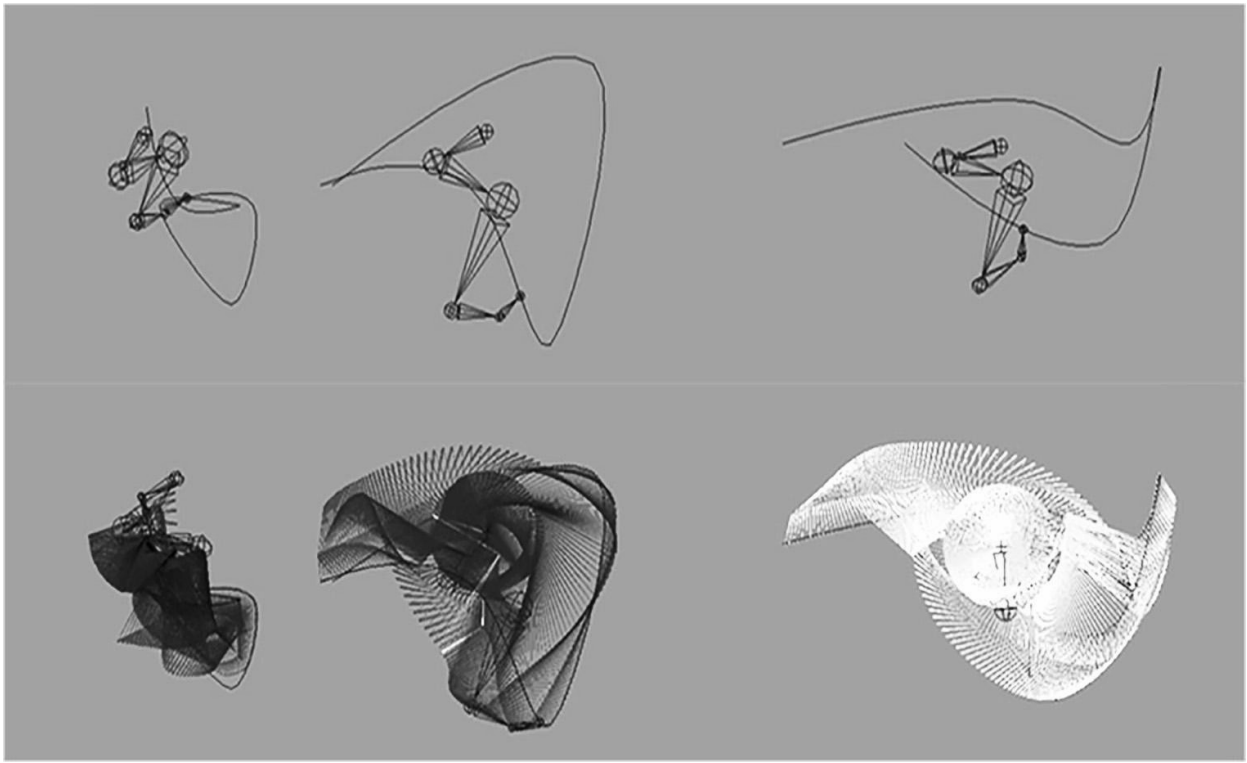

Figure 6. Emergent forms acquired via "long exposure" of sub-shapes connected to IK rigs

Figure 7. IK system combined with top level sweep animation

Multiple shapes can be combined in an IK system where their spatial relationships can be animated via affine transforms, informed by the IK effector. The total animated group can itself be constrained to animation paths, creating adaptive forms via the snapshot algorithm.

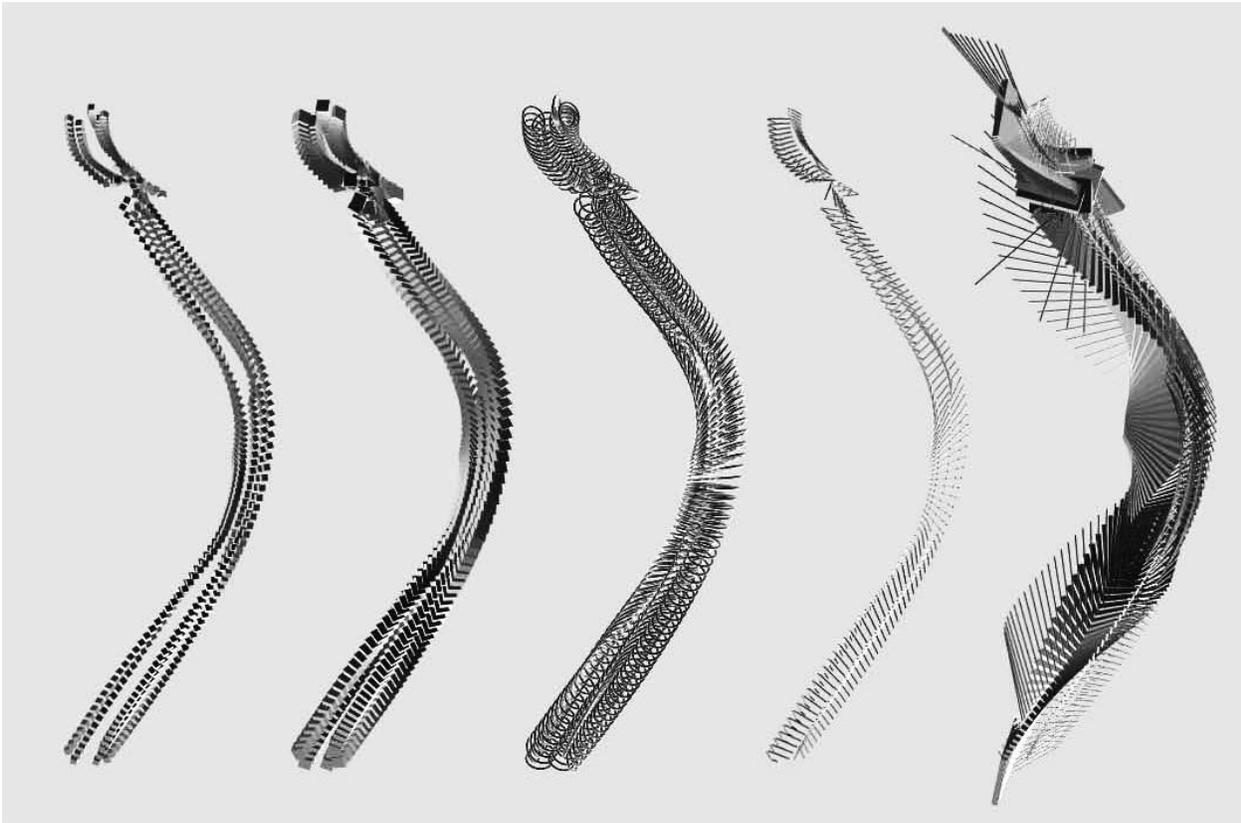

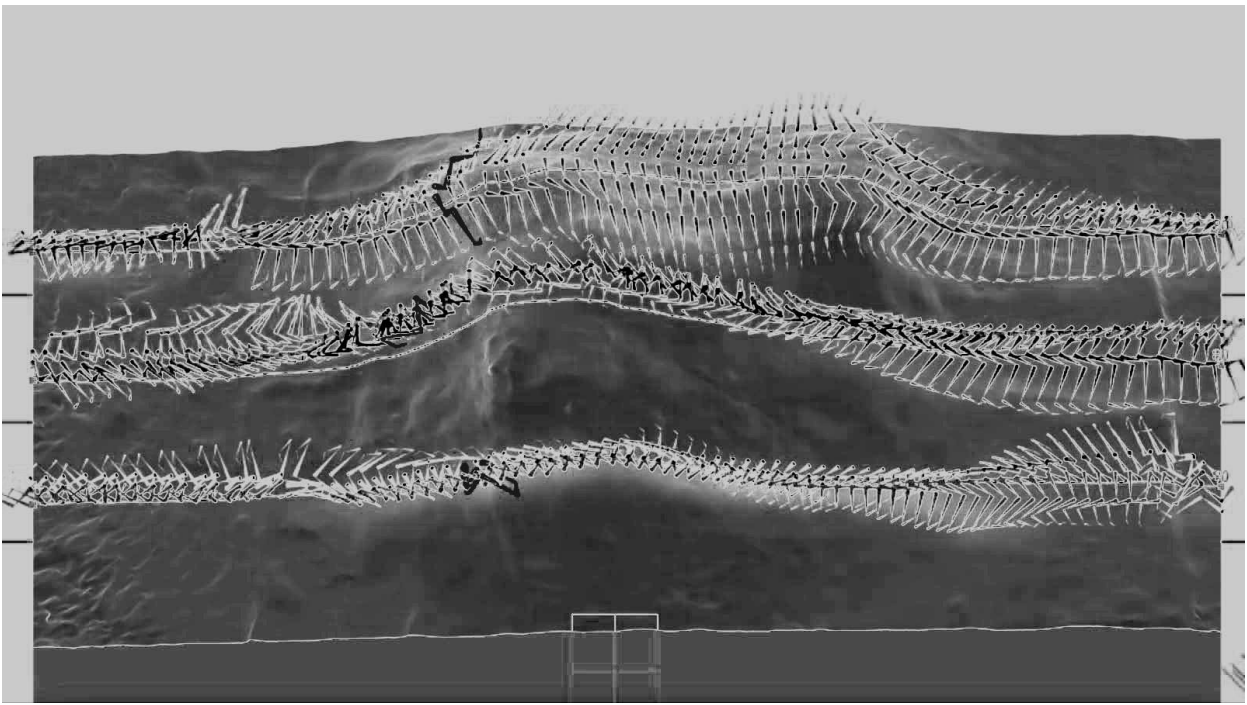

### 2.3. Trial III

In the third trial, swarm intelligence (SI) was investigated through the flight node of the MASH plug-in within Maya. The flight node is based on a particle system based on three basic rules of swarm behavior: collision avoidance, velocity matching and flock centering [12]. Additionally attractors or detractors can be placed and weighted in order to shift the position for flock centering, practically attracting or scaring the particles. An attractor was animated, based on the raster data, future work may include a detractor.

The landscape either literal, or as a fitness landscape is used as the environment for problem solving. Our examples were generated on a scanned 3d landscape model. Similar to the dog and frisbee metaphor put forward by Lynn, where a solution is chased in the fitness landscape by an agent [13]. In our system however, the intuition of a metaphorical dog is replaced by the collective intelligence of the swarm and the fitness landscape is replaced by a virtual one.

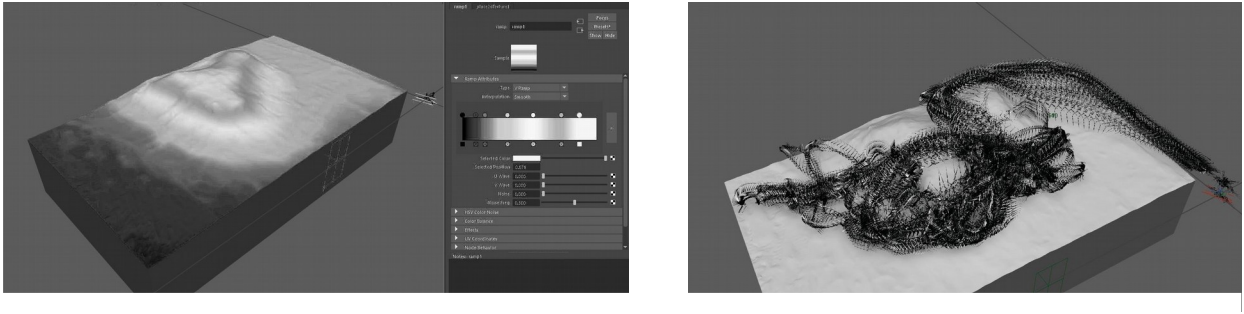

Figure 9. Adaptive forms generated via long exposure of the swarm animation on the informed landscape model

Agents were constrained to the terrain geometry, while the attractor was positioned below. This in effect created a situation where climbing was not favorable due to increased distance from the attractor. Multiple simulations were created with variations on the attractor animation and SI parameters.

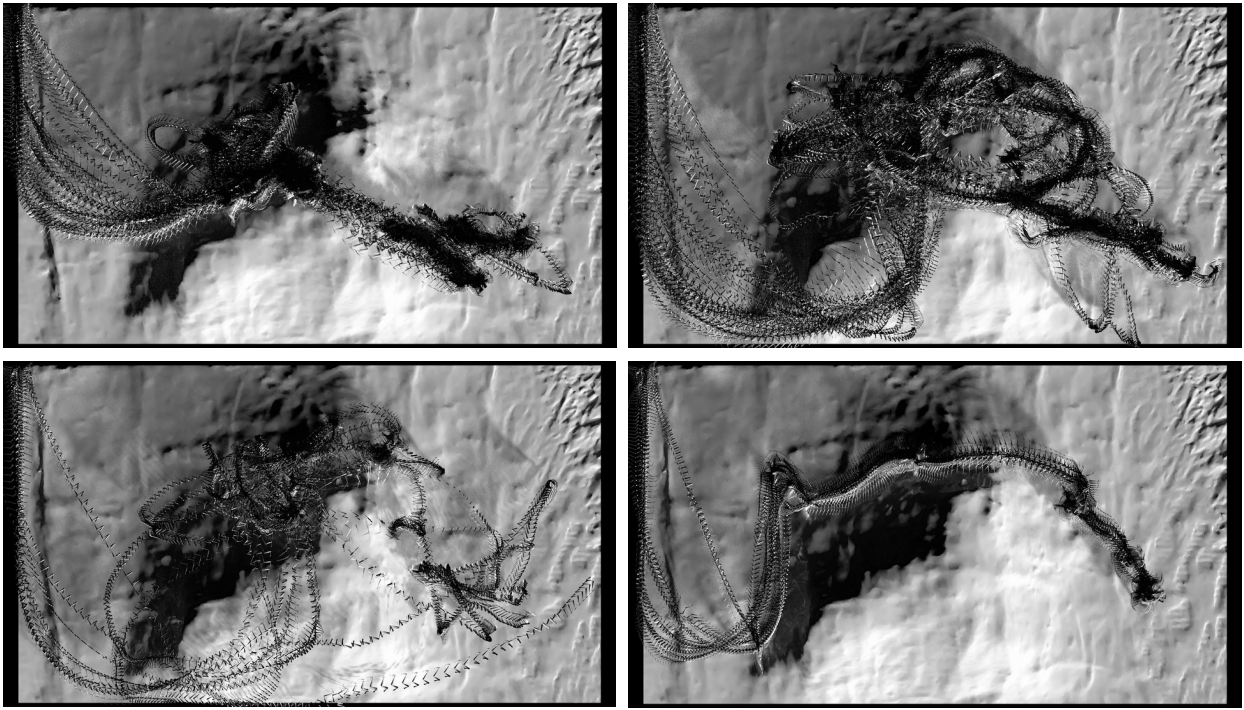

Figure 10. Varying results of the swarm test

### 3. RESULTS & DISCUSSION

Through the application of our trials, a form creation algorithm was created. Utilizing basic animation techniques and simple models, forms of a more complex nature can rapidly be generated. The results are based on the underlying landscape and the data field as well as user input. Manipulation of different parameters creates a situation where the designer has ample control on the results, similar in concept to the sweeping motions of a watercolor brush. The forms, while following the user motion, are shaped by the underlying data. By the application of these algorithms on scanned geometry, user input and collected data, complex, informed and adaptive forms can be generated.

The results are achieved through multi-layered animations of a set number of agents, each animated by the motion of their parts as well as on the landscape while collectively seeking the ideal path to the attractor, creating forms, adaptive to the terrain as well as the path of the attractor.

Solid models are acquired through the process, displaying characteristics of the embedded data fields, the scanned physical model and the user input through attractors. Resulting shapes although complex, include no ambiguity, similar results are achieved when the simulation is run with the same user input and data, unless parameters are intentionally randomized. With regards to the definitions made by De Wolf and Holvoet (2004), trial I can be considered as a simple parametric system that involves neither emergence nor self-organization since the system is always at an ordered state. In a cases where object pivots were offset however; the system did display emergent behavior which can be the subject of a further study.

Trial II contains cues of emergence, since resulting shapes do contain a "gestalt", not visible within the individual inputs. The system, with a static level of order does not exhibit self-organizational behavior.

It is when both the landscape data, IK animation and SI are combined in trial III that the system displays self-organization as well as emergence.

Despite containing no ambiguity, the system produces resulting forms that are irreducible. When agents are released upon the virtual landscape, the initial state of chaos evolves into higher order without external organizational input.

Our methods can be utilized for early design process of architectural design projects where complex field data is available or can be generated/ mined by the designer. On a different level, these algorithms can be used on analogue concept models in order to derive computation based results which can also be used for further stages of design. While our current work displays three dimensional forms of bird-like nature shaped by the wind data, it is possible to replace both the data input, and animation characteristics in order to generate designs of differing nature and varying levels of abstraction. Yet such possibilities were out of the timetable of our project schedule; it is our aim to introduce more variety in the future.

The main objective of our study was to bridge the design gap between material and digital processes. This condition is partially satisfied. While the process is fluent in carrying the form of the analogue model to the virtual space, most design input is through animating attractors digitally, and adjusting floating parameters. Future work may see these inputs connected to more immersive interfaces, in order to keep the process more fluent for a broader user base.

The main output of the work in its current form is visual, applied on the physical model via projection mapping. While such visual output is fast and suitable for quick design feedback, there may be potential for cases where the output can actually reshape the physical model through rapid additive or subtractive manufacturing processes.

## 4. CONCLUSION

Through this study, we aimed at combining physical models with generative systems in order to create processes for early design exploration. Forms are to be generated through the use of data fields, scanned models or actual terrain and user input of analog and parametric nature. Ambiguity is limited to user preference and not integral to the processes.

Based on a series of three trials, we have created a set of algorithms for the generation of informed geometries. The shaping factors can be as simple as curves, drawn in 2d space to multi-dimensional field data, utilized as rasters. Resulting shapes are complex, yet consistent and the system developed to generate them can be self-organizing, emergent or both.

Our work focuses on creating a fluent design process where the transition between the physical and the digital is as smooth as possible. Further work can be carried out in order to create a more physical interface for input from the designer while also utilizing the systems for digital fabrication techniques in order to reshape the original physical material via the output.

Results of our processes resemble the visual characteristics of avian creatures, combined with the wind field data, yet our system is open ended in nature where both the avian IK animation and the field data can be replaced with an endless range of material for differing contexts. We hope to test such possibilities in the future.